\documentclass[12pt]{iopart}

\usepackage{graphicx}
\begin{document}

\title[Simulation of jet quenching at RHIC and LHC]{Simulation of jet
quenching at RHIC and LHC}

\author{I P Lokhtin, A M Snigirev}

\address{Skobeltsyn Institute of Nuclear Physics, Moscow State University, 
Moscow, Russia}
\ead{Igor.Lokhtin@cern.ch}
\begin{abstract}
The model to simulate jet quenching effect in ultrarelativistic heavy ion 
collisions is presented. The model is the fast Monte-Carlo tool implemented to 
modify a standard PYTHIA jet event. The model has been generalized to the case 
of the ``full'' heavy ion event (the superposition of soft, hydro-type state 
and hard multi-jets) using a simple and fast simulation procedure for soft 
particle production. The model is capable of reproducing main features of
the jet quenching pattern at RHIC and is applyed to analyze novel jet quenching
features at LHC.
\end{abstract}

\pacs{25.75.-q, 12.38.Mh, 24.85.+p}

\section{Introduction:}

One of the important tools for studying the properties of quark-gluon plasma 
(QGP) in ultrarelativistic heavy ion collisions is the analysis of a QCD jet 
production. The medium-induced energy loss of energetic partons, ``jet 
quenching'', should be very different in the cold nuclear matter and QGP, 
resulting in many observable phenomena~\cite{baier_rev}. Recent RHIC data on 
high-p$_T$ particle production at $\sqrt{s}=200~A$ GeV are in agreement with 
the jet quenching hypothesis~\cite{Wang:2004}. At LHC, a new regime of heavy 
ion physics will be reached at 
$\sqrt{s_{\rm NN}}=5.5 A$ TeV where hard and semi-hard particle production can 
stand out against the underlying soft events. The initial gluon densities in 
Pb+Pb reactions at LHC are expected to be much higher than those at RHIC, 
implying a stronger partonic energy loss, observable in new channels.

In the most of available Monte-Carlo heavy ion event generators the 
medium-induced partonic rescattering and energy loss are either ignored or 
implemented insufficiently. Thus, in order to analyze RHIC data on high-p$_T$ 
hadron production and to test the sensitivity of LHC observables to the QGP 
formation, the development of adequate and fast Monte-Carlo tool to simulate 
the jet quenching is necessary.  

\section{Physics model and simulation procedure} 

The detailed description of physics model can be found in our recent 
paper~\cite{lokhtin-model}. The approach bases on an accumulating energy loss, 
the gluon radiation being associated with each parton scattering in the 
longitudinally expanding quark-gluon fluid and includes the interference effect using the modified 
radiation spectrum $dE/dl$ as a function of decreasing temperature $T$. The 
basic kinetic integral equation for the energy loss $\Delta E$ as a function of 
initial energy $E$ and path length $L$ has the form 
\begin{equation} 
\label{elos_kin}
\Delta E (L,E) = \int\limits_0^L dl \frac{dP(l)}{dl}\lambda (l)
\frac{dE(l,E)}{dl},~~~~ 
\frac{dP(l)}{dl} = \frac{1}{\lambda (l)} \exp{(-l/\lambda (l))}, 
\end{equation} 
where $l$ is the current transverse coordinate of a parton, $dP/dl$ is the 
scattering probability density, $dE/dl$ is the energy loss per unit length, 
$\lambda$ is in-medium mean free path. The collisional loss in high-momentum 
transfer limit, radiative loss in BDMS approximation~\cite{baier}, simple 
Gaussian parameterization of gluon emission angle distribution and realistic 
nuclear geometry are used. The QGP proper formation time $\tau_0$ and 
initial temperature $T_0$ are model parameters. 
  
The model has been constructed as the Monte-Carlo event generator PYQUEN 
(PYthia QUENched) and is available via Internet~\cite{pyquen}. The 
event-by-event simulation procedure includes the generation of the initial 
parton spectra with PYTHIA~\cite{pythia} and production vertexes 
at given impact parameter, rescattering-by-rescattering simulation of the 
parton path length in a dense zone, radiative and collisional energy loss per 
rescattering, final hadronization with the Lund string model for hard partons 
and in-medium emitted gluons.

The full heavy ion event is simulated as a superposition of soft hydro-type 
state and hard multi-jets. The simple approximation~\cite{lokhtin-model} of 
hadronic liquid at ``freeze-out'' stage has been used to treat soft part of the 
event. Then the hard part of the event includes PYQUEN multi-jets generated 
according to the binomial distribution. The mean number of jets produced in AA 
events at a given impact parameter is a product of the number of binary NN 
sub-collisions and the integral cross section of hard process in pp 
collisions with the minimal transverse momentum transfer $p_T^{\rm min}$. The 
extended in such a way model has been also constructed as the fast Monte-Carlo 
event generator~\cite{hydjet}. Note that ideologically similar approximation has 
been developed in~\cite{hirano}. 

\section{Jet quenching at RHIC}

\begin{figure}[htbp]
\begin{minipage}{18pc}
\includegraphics[width=18pc]{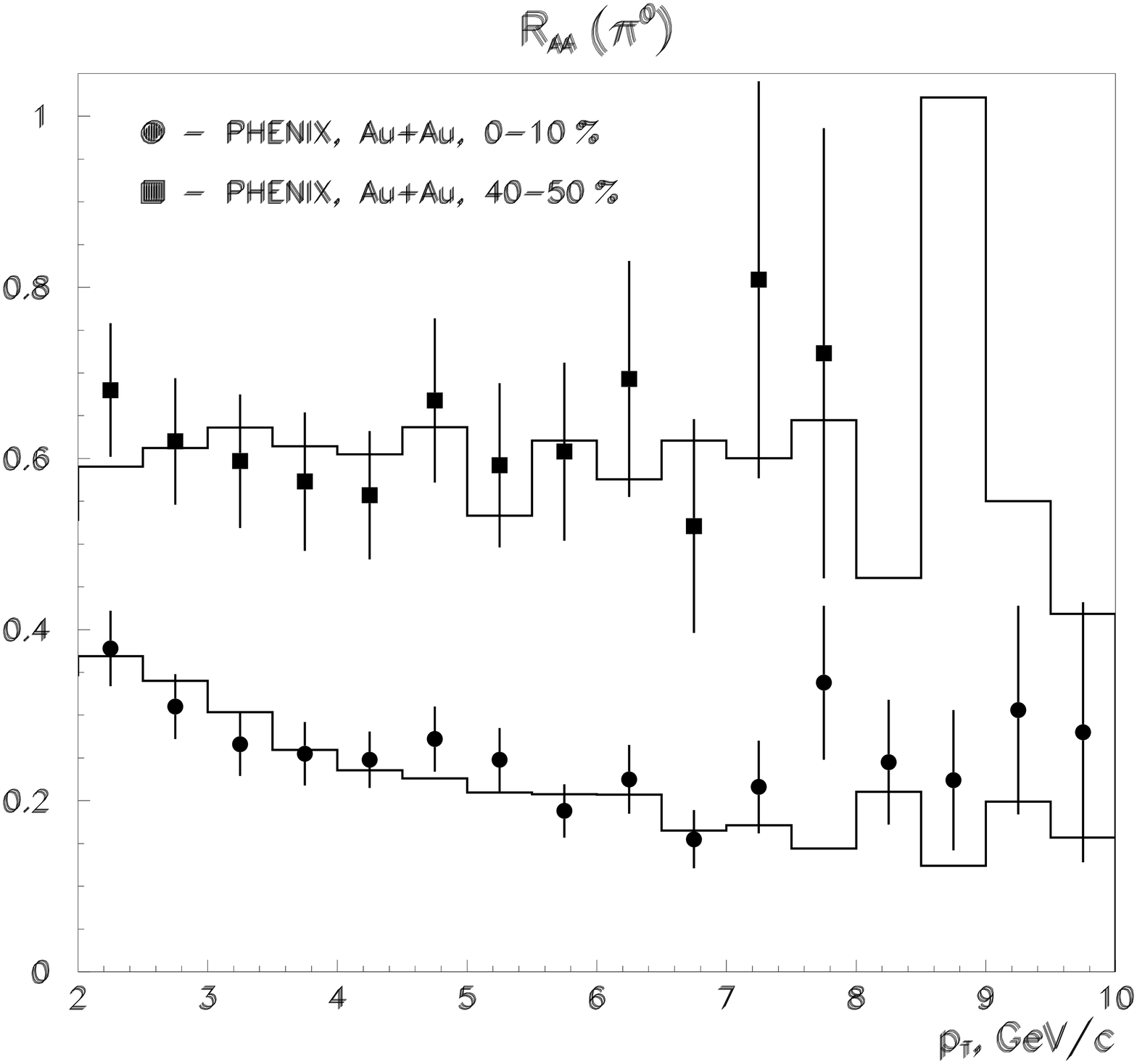}
\caption{The nuclear modification factor $R_{AA}$ for 
neutral pions in Au+Au collisions for two centrality sets. The points are  
PHENIX data, histograms are the model calculations. \label{rhic_rpt} }
\end{minipage}
\hspace{\fill}%
\begin{minipage}{18pc}
\includegraphics[width=18pc]{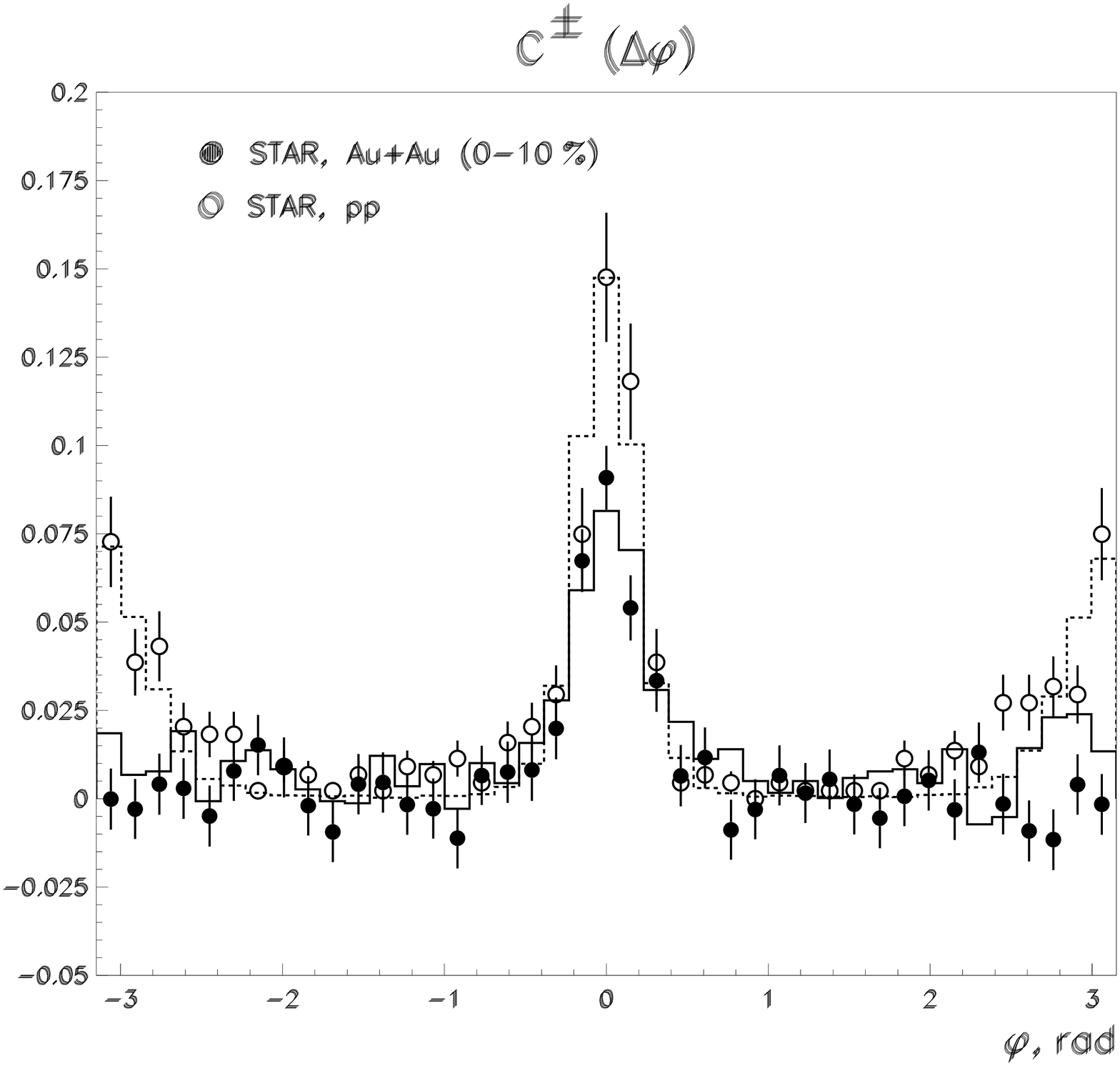}
\caption{The azimuthal two-particle correlation function 
for pp and for central Au+Au collisions. The points are STAR data, 
dashed and solid histograms are the model calculations for pp and Au+Au 
events respectively. \label{rhic_btb} }
\end{minipage} 
\end{figure}

In order to demonstrate the efficiency of the model, the jet quenching pattern 
in Au+Au collisions at RHIC was considered. Model parameters has been selected
from the fit of $\eta-$ and $p_T-$ spectra at RHIC for different event
centralities (see~\cite{lokhtin-model}
for details). In particular, the nuclear modification of the hardest domain 
of $p_T$-spectrum was used to extract initial QGP conditions: $T_0=500$ MeV 
and $\tau_ 0=0.4$ fm/$c$. Figure \ref{rhic_rpt} shows that our model 
reproduce $p_T-$ and centrality dependences of nuclear modification factor 
$R_{AA}$ (determined as a ratio of particle yields in $AA$ and pp 
collisions normalized on the number of binary NN sub-collisions) 
for $\pi_0$'s measured by PHENIX~\cite{phenix} quite well.
Another important tool to verify jet quenching is two-particle azimuthal 
correlation function $C(\Delta \varphi)$ -- the distribution over an azimuthal 
angle of high-$p_T$ hadrons in the event with $2$ GeV/$c<p_T<p_T^{\rm trig}$  
relative to that for the hardest ``trigger'' particle with $p_T^{\rm trig}>4$ 
GeV/$c$. Figure \ref{rhic_btb} presents $C(\Delta \varphi)$ in pp and in 
central Au+Au collisions (data from STAR~\cite{star}). Clear peaks in pp collisions at 
$\Delta \varphi = 0$ and $\Delta \varphi =\pi$ indicate a typical dijet event 
topology. However, for central Au+Au collisions the peak near $\pi$ disappears. 
It can be interpreted as the observation of monojet events due to the 
absorption of one of the jets in a dense medium. Figure \ref{rhic_btb} demonstrates that 
measured suppression of azimuthal back-to-back correlations is well reproduced 
by our model.  

We leave beyond the scope of this paper the analysis of such important RHIC 
observables as the azimuthal anisotropy and particle ratios, which are 
sensitive to the soft physics. In order to study them, a more careful treatment 
of low-$p_T$ particle production than our simple approach is 
needed~\cite{Amelin:2006qe} 
(the detailed description of space-time structure of freeze-out region, 
resonance decays, etc.). 

\section{Jet quenching at LHC}

The developed model was applied to analyze various novel features of jet 
quenching at the LHC (see~\cite{lokhtin-jff,lokhtin-bjet,lokhtin-mujet} for 
details). Let us just to enumerate some main issues. 

{\em Jets tagged by leading hadrons.} The relation between in-medium softening 
jet fragmentation function (measured with leading hadrons) and suppression 
of jet rates due to energy loss out of jet cone, was analyzed in the  
work~\cite{lokhtin-jff}. The specific anti-correlation 
between two effects allows one to probe parton energy loss mechanism
(small-angular radiation versus wide-angular radiation and collisional loss). 

{\em Jets induced by heavy quarks.} The possibility to observe B-jets  
tagged by an energetic muon in 
heavy ion collisions was analyzed in the work~\cite{lokhtin-bjet}. The 
significant softening b-jet fragmentation function due to b-quark energy loss 
is predicted. 

{\em Z/$\gamma^*$+jet production.}  The channel with dimuon tagged jet 
production in heavy ion collisions was analyzed in the 
work~\cite{lokhtin-mujet}. We have found that the medium-induced partonic 
energy loss can result in significant $P_T$-imbalance between $\mu ^+\mu ^-$ 
pair and a leading particle in a jet, which is quite visible even for moderate 
loss. 

\section{Conclusions} 

The method to simulate jet quenching in heavy ion collisions has been 
developed. The model is the fast Monte-Carlo tool implemented to modify 
a standard PYTHIA jet event. The full heavy ion event is obtained as a 
superposition of a soft hydro-type state and hard multi-jets. The model is 
capable of reproducing main features of the jet quenching pattern at RHIC 
(the $p_T-$dependence of the nuclear modification factor and the suppression of 
azimuthal back-to-back correlations). The model was also applied to probe jet 
quenching in various new channels at LHC energy: jets tagged by leading 
particles, b-jets, dilepton-jet correlations. The further development of the 
model focusing on a more detailed description of low-$p_T$ particle production 
is in the progress. 

\section*{Acknowledgments}

Discussions with A.~Morsch, L.I.~Malinina, C.~Roland, L.I.~Sarycheva and 
B.~Wyslouch are gratefully acknowledged. I.L. thanks the organizers 
of ``Quark Matter 2006'' for the warm welcome and hospitality. This work is 
supported by grants N 04-02-16333 and 06-02-27355 of Russian Foundation for 
Basic Research and Contract N 02.434.11.7074 of Russian Ministry of Science 
and Education.

\end{document}